\documentclass[conference]{IEEEtran}

\usepackage{amsmath,amssymb}
\usepackage{booktabs}
\usepackage{array}
\usepackage{longtable}
\usepackage{pdflscape}
\usepackage{graphicx}
\usepackage{algorithm}
\usepackage{algpseudocode}
\usepackage{multirow}
\usepackage{url}
\usepackage{cite}
\usepackage[hidelinks]{hyperref}
\usepackage{balance}

\newcommand{\Baseline}{\textsc{Baseline}}
\newcommand{\Liberal}{\textsc{Liberal}}

\title{LLMs as Teaching Assistants for Mathematics Exam Grading: Reliability, and Practical Usability}

\author{Aastha Sapkota and M. G. Sarwar Murshed}

\author{%
  \IEEEauthorblockN{Aastha~Sapkota and M.~G.~Sarwar Murshed}
  \IEEEauthorblockA{\textit{Computer Science Department} \\
  \textit{University of Wisconsin - Green Bay, Green Bay, WI, USA}}
}

\begin{document}
\maketitle

\begin{abstract}
Open-ended mathematics exams are valuable because they assess reasoning, proof construction, algorithmic thinking, and communication of intermediate steps. They are also difficult to grade at scale because instructors must apply partial-credit rubrics consistently while giving feedback that helps students repair misconceptions. This paper evaluates six contemporary large language model (LLM) configurations, Gemini 3.1 Pro Extended, Gemini 3.5 Flash, ChatGPT 5.5 Pro Extended, ChatGPT 5.5 Thinking, Claude Pro Opus 4.7, and Claude Sonnet 4.6, as grading assistants for an undergraduate discrete mathematics examination. The study compares two grading policies. The \Baseline{} policy uses a stricter rubric-following prompt that emphasizes explicit evidence and complete justification. The \Liberal{} policy was added after preliminary grading showed that the baseline condition sometimes applied harsh point deductions and failed to recognize valid partial reasoning. Agreement with human grading is measured at both the question and exam-total levels using mean absolute error, root mean squared error, normalized root mean squared error, Pearson correlation, and exact agreement. The results show that liberal partial-credit prompting reduces average question-level error for every evaluated model family. ChatGPT 5.5 Thinking (\Liberal{}) has the lowest average question-level MAE (1.87) and RMSE (2.53), while Gemini 3.1 Pro Extended (\Liberal{}) has the lowest total-score MAE (8.00) and RMSE (10.66). However, the strongest total-score Pearson correlation occurs under Gemini 3.1 Pro Extended (\Baseline{}) at 0.58, showing that point calibration and rank preservation remain distinct goals. We also report practical usability observations: ChatGPT completed the grading workflow in one attempt, Claude required continuation because of usage limits, and Gemini required dataset splitting after file-handling failures on the full dataset. These findings support a cautious deployment model in which LLMs act as auditable teaching assistants rather than autonomous graders.
\end{abstract}

\begin{IEEEkeywords}
Large language models, automated grading, mathematics education, short-answer grading, rubric-based assessment, feedback quality, human-in-the-loop AI, partial credit.
\end{IEEEkeywords}

\section{Introduction}
Assessment in mathematics and computer science education is not merely a mechanism for assigning scores. It is a central part of instruction: grading communicates what counts as valid reasoning, reveals misconceptions, and determines whether students receive timely opportunities to improve. Open-ended mathematics questions are especially important because they require students to show intermediate reasoning, construct proofs, interpret definitions, and apply algorithms. These same properties make grading difficult. A correct final answer may conceal invalid reasoning; a wrong final answer may deserve substantial partial credit; and semantically equivalent solutions may use different notation, organization, or explanatory style.

Large classes amplify these challenges. Human graders must apply rubrics repeatedly across heterogeneous responses while managing fatigue, time pressure, and borderline cases. Instructors often attempt to increase reliability through detailed rubrics, teaching-assistant calibration, and discussion of representative examples. Even with these safeguards, score variance and delayed feedback remain common in open-ended assessment.

Automated short-answer grading systems promise efficiency, but many earlier systems depend on surface similarity, hand-engineered features, or narrow answer templates. Recent LLMs can interpret natural language and mathematical explanations more flexibly, motivating their use as grading assistants for essays, STEM short answers, programming assignments, and mathematical reasoning tasks \cite{schneider2023towards,impey2025using,lee2024college,alkafaween2025automating,satcharattanachot2025automated,baral2024automated}. However, the central question is not whether an LLM can produce a plausible score. The more important question is whether it can support a grading process that is accurate, consistent, transparent, and pedagogically useful.

A deployable assistant must satisfy at least five requirements. First, it must agree with human grading at the point level, not only at the ranking level. Second, it must follow an instructor-specified rubric rather than replace the rubric with its own implicit preferences. Third, it must expose enough evidence and confidence information for instructors to audit its decisions. Fourth, it must generate feedback that is specific to the student's reasoning rather than generic praise or criticism. Fifth, it must be practically usable within real grading constraints, including file handling, context size, output stability, token limits, and grading speed.

This paper presents an empirical evaluation of LLMs as practical grading assistants for open-ended mathematics assessment. The evaluation covers six contemporary LLM configurations and distinguishes between a \Baseline{} condition and a \Liberal{} condition. The \Baseline{} condition is a stricter rubric-following prompt that requires explicit evidence and complete justification. The \Liberal{} condition was added after preliminary runs showed that the baseline condition sometimes graded too harshly; it tests whether a more generous partial-credit interpretation better matches human grading.

\subsection{Research Questions}
This paper is guided by five research questions.

\textbf{RQ1: Agreement.} How closely do LLM-generated scores match human scores on open-ended mathematics exam questions?

\textbf{RQ2: Reliability.} Do LLM graders preserve the ordering of students and produce stable score patterns across questions and total exam scores?

\textbf{RQ3: Policy sensitivity.} How much do scores change when the same model is evaluated under the stricter \Baseline{} policy versus the more permissive \Liberal{} partial-credit policy?

\textbf{RQ4: Feedback quality.} What output structure is required for LLM feedback to be usable by instructors and students in a human-in-the-loop grading workflow?

\textbf{RQ5: Practical usability.} How do file-handling behavior, dataset size, token limits, session limits, and response speed affect the feasibility of using LLMs as grading assistants?

\subsection{Contributions}
The paper makes the following contributions:
\begin{itemize}
    \item It presents a mathematics exam grading evaluation using seven open-ended discrete mathematics questions and instructor-assigned reference scores.
    \item It compares Gemini 3.1 Pro Extended, Gemini 3.5 Flash, ChatGPT 5.5 Pro Extended, ChatGPT 5.5 Thinking, Claude Pro Opus 4.7, and Claude Sonnet 4.6 under baseline and liberal grading policies.
    \item It reports question-level and total-score performance using MAE, RMSE, NRMSE, Pearson correlation, and exact agreement.
    \item It shows that liberal partial-credit prompting improves point error for all evaluated model families, but may not improve rank correlation in every case.
    \item It adds a practical usability analysis documenting successful one-pass grading, file-handling problems, dataset splitting, token-limit interruptions, and speed differences.
    \item It proposes a rubric-calibrated LLM Evaluation Assistant with algorithms for rubric normalization, policy-specific grading, cross-submission audit, structured feedback, and human approval.
\end{itemize}

\section{Related Work}
\subsection{Automated Grading Before LLMs}
Automated grading has a long history in education. Objective assessments such as multiple-choice questions and constrained numerical responses are comparatively straightforward because the answer space is limited. Open-ended short answers, proofs, essays, and code submissions are substantially harder because they require semantic interpretation and partial-credit judgment. Traditional automated short-answer grading systems have used lexical overlap, semantic similarity, dependency features, or task-specific classifiers. These approaches can be effective when reference answers are narrow, but they often struggle with alternative valid explanations, incomplete reasoning, and transfer across instructors or courses.

\subsection{LLM-Based Grading and Feedback}
LLMs have renewed interest in automated assessment because they can parse free-form responses, follow rubrics, and generate explanations. Schneider et al. study LLM-based autograding for short textual answers and frame LLMs as a useful second opinion rather than a complete replacement for human graders \cite{schneider2023towards}. Impey et al. evaluate LLMs for grading student science writing under different prompt and rubric configurations \cite{impey2025using}. Lee and Song propose a college exam grading system based on LLM AI models \cite{lee2024college}. In programming education, LLMs have been used not only to grade submissions but also to generate test suites for autograding \cite{alkafaween2025automating}. In STEM and mathematics education, recent work evaluates automated feedback for open-ended responses and emphasizes the importance of partial credit, explanation correctness, and feedback specificity \cite{satcharattanachot2025automated,baral2024automated,lee2026personalized}.

Despite these advances, LLM grading remains sensitive to prompts, rubrics, answer order, model behavior, and the surrounding workflow. Ferreira Mello et al. compare LLM-based short-answer grading with traditional models in the prompt-engineering era \cite{ferreira2025automatic}. Broader surveys of LLMs and LLM agents in education emphasize that instructional integration requires reliability, transparency, alignment with pedagogy, and appropriate human oversight rather than accuracy alone \cite{wang2026large,chu2025llmagentforeducation}.

\subsection{Human-in-the-Loop and Rubric-Aware Grading}
A key limitation of simple LLM graders is that they often grade each answer independently, without a memory of previous decisions or a mechanism to detect inconsistency. Human graders rarely work this way. They calibrate on sample responses, revise rubrics when unexpected answer patterns appear, compare borderline cases, and audit final scores. Recent LLM grading frameworks attempt to operationalize these practices. Chu et al. introduce human-in-the-loop automated grading and guideline optimization mechanisms \cite{chu2024llm,chu2025llm}. Xie et al. propose Grade-Like-a-Human, a multi-agent framework that refines rubrics using real answers, uses cross-item memory for consistency, and applies post-grading audit and feedback \cite{xie2025grade}. Our work adopts the same philosophy but focuses on undergraduate discrete mathematics exams with proof-oriented and algorithmic questions.

\section{Problem Formulation}
Let $Q = \{q_1, \ldots, q_J\}$ denote a set of exam questions and let $A = \{a_1, \ldots, a_N\}$ denote anonymized student submissions. For each student $i$ and question $j$, the human reference score is $y_{ij}$ and the model-produced score is $\hat{y}^{(m,p)}_{ij}$, where $m$ denotes the LLM configuration and $p \in \{\Baseline{}, \Liberal{}\}$ denotes the grading policy. Let
\begin{equation}
T_i = \sum_{j=1}^{J} y_{ij}
\end{equation}
and
\begin{equation}
\hat{T}^{(m,p)}_i = \sum_{j=1}^{J} \hat{y}^{(m,p)}_{ij}
\end{equation}
denote the human and LLM total exam scores.

The goal is not to optimize a single predictive metric. A useful grading assistant must satisfy multiple desiderata:
\begin{enumerate}
    \item \textbf{Point agreement:} $\hat{y}_{ij}$ should be close to $y_{ij}$ for each response.
    \item \textbf{Rank reliability:} if one student demonstrates stronger performance than another, the model should usually preserve that ordering.
    \item \textbf{Rubric faithfulness:} each score should be traceable to explicit rubric criteria.
    \item \textbf{Feedback quality:} generated feedback should explain awarded and missing credit in a way that supports learning.
    \item \textbf{Auditability:} the system should flag uncertain, inconsistent, or operationally incomplete cases for human review.
    \item \textbf{Operational feasibility:} the model and interface should handle the dataset without excessive manual splitting, repeated continuation, or unrecoverable file failures.
\end{enumerate}

We therefore treat LLM grading as a decision-support problem. The model proposes scores and feedback; the instructor remains responsible for final grading decisions.

\section{Rubric-Calibrated LLM Evaluation Assistant}
\subsection{System Overview}
\begin{figure}[t]
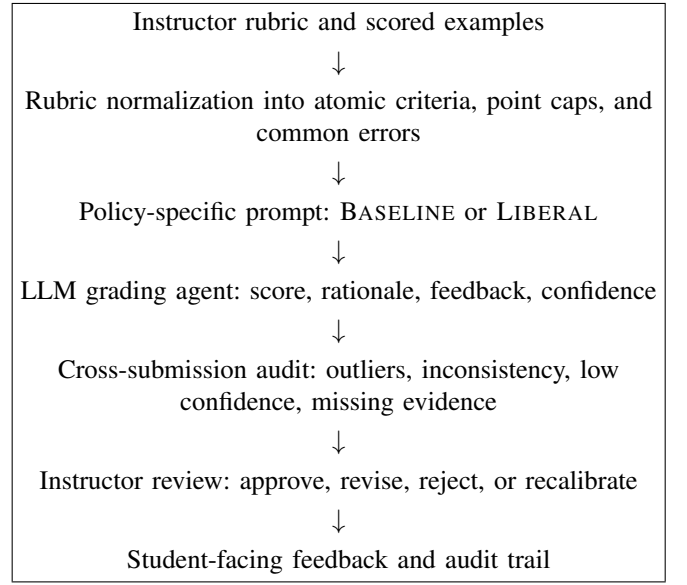

\centering
\fbox{\begin{minipage}{0.95\linewidth}
\centering
Instructor rubric and scored examples\\[1mm]
$\downarrow$\\[1mm]
Rubric normalization into atomic criteria, point caps, and common errors\\[1mm]
$\downarrow$\\[1mm]
Policy-specific prompt: \Baseline{} or \Liberal{}\\[1mm]
$\downarrow$\\[1mm]
LLM grading agent: score, rationale, feedback, confidence\\[1mm]
$\downarrow$\\[1mm]
Cross-submission audit: outliers, inconsistency, low confidence, missing evidence\\[1mm]
$\downarrow$\\[1mm]
Instructor review: approve, revise, reject, or recalibrate\\[1mm]
$\downarrow$\\[1mm]
Student-facing feedback and audit trail
\end{minipage}}
\caption{Proposed rubric-calibrated human-in-the-loop grading workflow. The empirical study evaluates the grading outputs; the deployment framework adds calibration, audit, feedback control, and instructor approval.}
\label{fig:workflow}
\end{figure}

Fig.~\ref{fig:workflow} presents the proposed LLM Evaluation Assistant. The workflow is designed to resemble a careful human grading process. It begins with instructor-provided exam questions, rubrics, and optionally a small set of previously graded examples. The rubric is normalized into atomic criteria so that the LLM can assign and explain partial credit. The model then grades each submission under a specified grading policy. A cross-submission audit module compares model decisions across students and flags inconsistent, low-confidence, or outlier scores. The instructor reviews flagged cases and can approve, revise, or reject model outputs. The final output is an auditable package containing scores, criterion-level rationales, student-facing feedback, confidence estimates, and an instructor-visible audit trail.

\subsection{Baseline and Liberal Grading Policies}
The study uses two grading policies because mathematics grading often requires a balance between formal correctness and recognition of partial reasoning.

The \Baseline{} policy is the stricter rubric-following condition. In practice, this condition behaved as a stricter grading mode: it penalized missing justifications, ambiguous notation, unsupported claims, and incorrect intermediate steps. This behavior is appropriate when an exam emphasizes proof rigor or when the rubric explicitly requires formal derivation. However, preliminary review showed that the baseline prompt could also become overly harsh by withholding credit for responses that demonstrated partially correct mathematical ideas.

The \Liberal{} policy was added to address this limitation. It awards credit for mathematically valid ideas even when notation, wording, or final presentation is imperfect. It is appropriate when the instructor wishes to reward conceptual understanding and partial progress. The two policies do not represent different ground truths. They represent different operational interpretations of partial credit. Comparing them allows us to measure prompt-policy sensitivity and determine whether each model aligns better with human graders under a conservative or permissive interpretation of the rubric.

\subsection{Algorithmic Workflow}
Algorithm~\ref{alg:grading} describes the grading workflow. Algorithm~\ref{alg:audit} describes the post-grading audit mechanism, which is inspired by the human practice of comparing borderline answers after an initial pass. The audit algorithm does not replace the instructor. It creates a prioritized review queue.

\begin{algorithm}[t]
\caption{Rubric-Calibrated LLM Grading}
\label{alg:grading}
\begin{algorithmic}[1]
\Require Questions $Q$, submissions $A$, instructor rubrics $R$, optional scored examples $E$, model $m$, policy $p \in \{\Baseline{}, \Liberal{}\}$
\Ensure Candidate scores, rationales, feedback, confidence, and audit flags
\For{each question $q_j \in Q$}
    \State $\tilde{R}_j \gets \textsc{NormalizeRubric}(q_j, R_j, E_j)$
    \State $P_j \gets \textsc{BuildPrompt}(q_j, \tilde{R}_j, p)$
    \For{each student answer $a_{ij} \in A$}
        \State $o_{ij} \gets \textsc{LLMGrade}(m, P_j, a_{ij})$
        \State Parse $o_{ij}$ into score $\hat{y}_{ij}$, criterion evidence, feedback, and confidence
        \State Enforce score range and rubric point caps
    \EndFor
    \State $F_j \gets \textsc{Audit}(q_j, \tilde{R}_j, \{(a_{ij},\hat{y}_{ij},o_{ij})\}_{i=1}^N)$
\EndFor
\State \Return all candidate grades and instructor review queue $\cup_j F_j$
\end{algorithmic}
\end{algorithm}

\begin{algorithm}[t]
\caption{Cross-Submission Audit and Review Queue Construction}
\label{alg:audit}
\begin{algorithmic}[1]
\Require Question $q$, normalized rubric $\tilde{R}$, graded answer set $D=\{(a_i,\hat{y}_i,o_i)\}_{i=1}^N$, group size $c$, regrouping rounds $B$
\Ensure Flagged cases $F$
\State $F \gets \emptyset$
\For{$b = 1$ to $B$}
    \State Partition $D$ into random groups $G_1,\ldots,G_K$ of size at most $c$
    \For{each group $G_k$}
        \State Ask an audit agent to compare answers, scores, rationales, and rubric criteria within $G_k$
        \State Flag score-rubric mismatches, inconsistent treatment of similar reasoning, low confidence, unsupported feedback, and incomplete outputs
        \State Add flagged cases to $F$ with explanations and suggested review priority
    \EndFor
\EndFor
\State Merge duplicate flags and rank by severity, score impact, and confidence
\State \Return $F$
\end{algorithmic}
\end{algorithm}

\section{Dataset and Experimental Design}
\subsection{Mathematics Exam Dataset}
The empirical study uses an undergraduate discrete mathematics examination with seven open-ended questions, denoted Q1--Q7. The questions cover standard topics in discrete mathematics and algorithmic reasoning, including sets and Cartesian products, proof reasoning, asymptotic complexity, the Euclidean algorithm, probability, and shortest-path algorithms. Each student response was graded by human graders, producing both per-question scores and a total exam score. The LLM scores are compared against these human scores as the reference standard.

The dataset is intentionally different from many short-answer benchmarks. The questions require mathematical reasoning rather than simple fact recall, and partial credit depends on intermediate steps. This makes the setting useful for evaluating whether LLMs can assist instructors in grading responses that usually require human judgment.

\subsection{Models and Prompting Conditions}
The study evaluates six LLM configurations: Gemini 3.1 Pro Extended, Gemini 3.5 Flash, ChatGPT 5.5 Pro Extended, ChatGPT 5.5 Thinking, Claude Pro Opus 4.7, and Claude Sonnet 4.6. Each model is evaluated under \Baseline{} and \Liberal{} policies. The same exam questions, rubrics, and grading instructions are used across model-policy pairs to improve comparability. Each model produces a per-question score for each student response, and an exam-total score is obtained from the question scores.

The table reports $N$ for each row because the number of usable scored responses differs in a few cases. Most conditions include $N=28$ submissions. ChatGPT 5.5 Pro Extended in the baseline condition reports $N=27$, and Claude Sonnet 4.6 reports $N=27$ for Q7 in both policies. These row-level sample sizes are retained in the results rather than imputed.

\subsection{Practical Execution Notes}
The experiment also recorded practical execution observations. ChatGPT completed the grading task in one attempt. Claude Pro Opus 4.7 reached usage or session limits and required a continuation action before grading could proceed. Gemini showed file-handling problems, including cases where the dataset file was not found. Gemini 3.5 Flash did not reliably handle the full dataset in a single run: initial attempts on the large dataset produced inaccurate or unacceptable output, so the dataset was split into three smaller parts before final scoring. Gemini was fast in both basic and pro model conditions, while Claude Pro Opus 4.7 was the slowest observed condition. These observations are not a controlled latency benchmark; they are deployment-relevant notes about workflow friction that would matter to instructors using commercial LLM interfaces.

\subsection{Evaluation Metrics}
We evaluate agreement using five metrics. Let $y_i$ denote a human score and $\hat{y}_i$ denote a model score for a fixed question or for the total exam.

Mean Absolute Error (MAE) measures average point deviation:
\begin{equation}
\mathrm{MAE} = \frac{1}{N}\sum_{i=1}^{N}|\hat{y}_i-y_i|.
\end{equation}

Root Mean Squared Error (RMSE) penalizes larger deviations more strongly:
\begin{equation}
\mathrm{RMSE}=\sqrt{\frac{1}{N}\sum_{i=1}^{N}(\hat{y}_i-y_i)^2}.
\end{equation}

Normalized RMSE (NRMSE) divides RMSE by the observed human-score range, enabling comparison across questions with different point scales:
\begin{equation}
\mathrm{NRMSE}=\frac{\mathrm{RMSE}}{\max_i y_i-\min_i y_i}.
\end{equation}

Pearson correlation measures rank-preserving linear association:
\begin{equation}
r=\frac{\sum_{i=1}^{N}(\hat{y}_i-\bar{\hat{y}})(y_i-\bar{y})}{\sqrt{\sum_{i=1}^{N}(\hat{y}_i-\bar{\hat{y}})^2}\sqrt{\sum_{i=1}^{N}(y_i-\bar{y})^2}}.
\end{equation}

Exact agreement is the fraction of responses for which the model score exactly matches the human score:
\begin{equation}
\mathrm{Exact}=\frac{1}{N}\sum_{i=1}^{N}\mathbf{1}(\hat{y}_i=y_i).
\end{equation}

We report question-level metrics for each question and model-policy pair. The total-score row is computed directly from human total scores and model total scores, not by averaging question-level metrics.

\section{Results}
\subsection{Question-Level and Total-Score Agreement}
\begin{figure}[t]
\centering
\includegraphics[width=\linewidth]{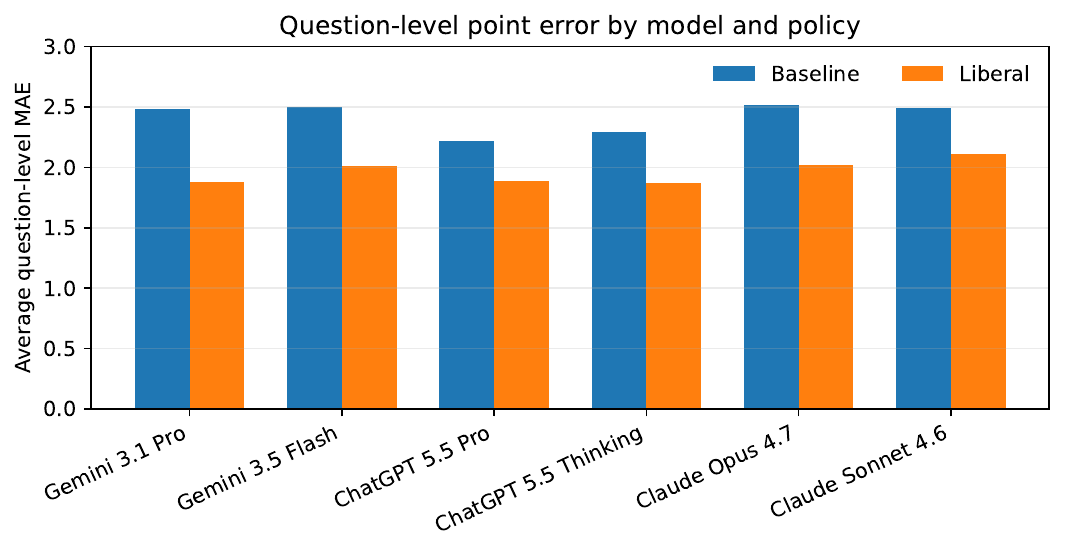}
\caption{Average question-level MAE by model and grading policy. Lower values indicate better point calibration.}
\label{fig:question-mae}
\end{figure}

\begin{figure}[t]
\centering
\includegraphics[width=\linewidth]{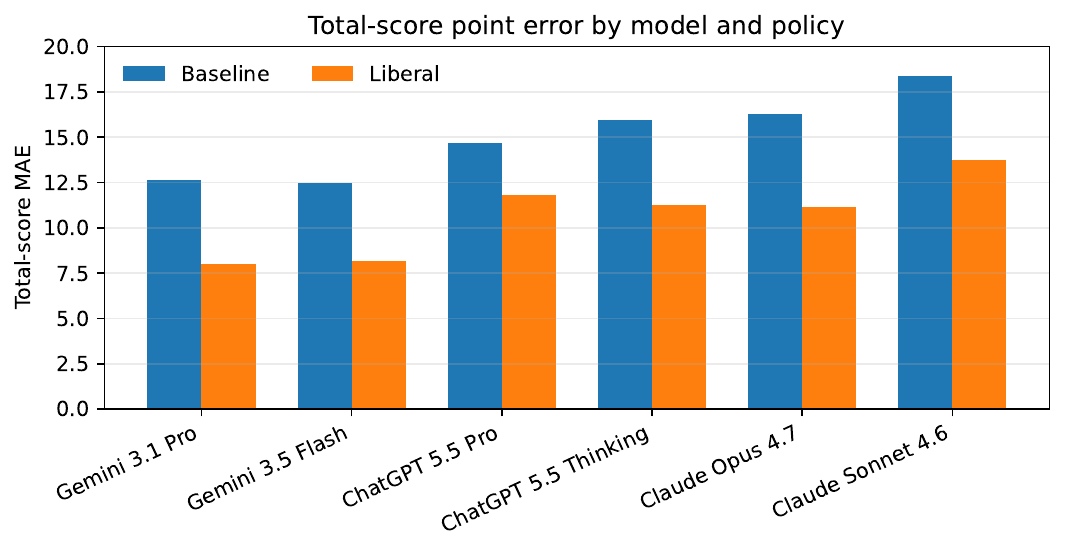}
\caption{Total-score MAE by model and grading policy. The total-score row is computed from exam totals rather than averaged from question-level metrics.}
\label{fig:total-mae}
\end{figure}

Table~\ref{tab:llm-grading-performance} reports the full grading performance for all models, policies, and questions. Table~\ref{tab:llm-grading-averages} summarizes average question-level performance across Q1--Q7. Fig.~\ref{fig:question-mae} and Fig.~\ref{fig:total-mae} visualize point-error patterns for the question-level and total-score settings.

\subsection{Main Findings}
\subsubsection{Liberal evaluation improves point calibration for every model family}
The most consistent result is that the \Liberal{} policy improves point-error metrics for all evaluated model families. At the question level, Gemini 3.1 Pro Extended improves from 2.48 to 1.88 MAE, Gemini 3.5 Flash improves from 2.50 to 2.01, ChatGPT 5.5 Pro Extended improves from 2.22 to 1.89, ChatGPT 5.5 Thinking improves from 2.29 to 1.87, Claude Pro Opus 4.7 improves from 2.52 to 2.02, and Claude Sonnet 4.6 improves from 2.49 to 2.11. This pattern supports the decision to add liberal evaluation after observing that the baseline prompt sometimes graded too strictly.

The lowest average question-level MAE is achieved by ChatGPT 5.5 Thinking (\Liberal{}) at 1.87. Gemini 3.1 Pro Extended (\Liberal{}) is nearly tied at 1.88, and ChatGPT 5.5 Pro Extended (\Liberal{}) follows at 1.89. ChatGPT 5.5 Thinking (\Liberal{}) also has the lowest average question-level RMSE at 2.53. Gemini 3.1 Pro Extended (\Liberal{}) and ChatGPT 5.5 Thinking (\Liberal{}) tie for the best average NRMSE at 0.36.

\subsubsection{Total-score error also improves under liberal grading}
The same pattern appears at the total-score level. Gemini 3.1 Pro Extended improves from 12.66 to 8.00 total-score MAE under the liberal policy. Gemini 3.5 Flash improves from 12.46 to 8.16. ChatGPT 5.5 Pro Extended improves from 14.70 to 11.79. ChatGPT 5.5 Thinking improves from 15.96 to 11.25. Claude Pro Opus 4.7 improves from 16.29 to 11.14. Claude Sonnet 4.6 improves from 18.39 to 13.75. The best total-score MAE is Gemini 3.1 Pro Extended (\Liberal{}) at 8.00, followed closely by Gemini 3.5 Flash (\Liberal{}) at 8.16.

\subsubsection{Point calibration and rank preservation diverge}
Although liberal prompting improves point error, it does not always improve Pearson correlation. At the total-score level, Gemini 3.1 Pro Extended (\Baseline{}) has the strongest Pearson correlation at 0.58, while the same model under \Liberal{} has lower total-score correlation at 0.34. Similarly, the total-score Pearson correlation for Claude Pro Opus 4.7 changes only slightly from 0.48 to 0.46 despite a large improvement in MAE. These results reinforce a central principle: a model can be better calibrated in points while not necessarily preserving ranking better, and a model can preserve ranking while assigning poorly calibrated scores.

At the question level, ChatGPT 5.5 Pro Extended (\Liberal{}) has the highest average Pearson correlation at 0.56, with ChatGPT 5.5 Pro Extended (\Baseline{}) close behind at 0.55. Gemini 3.5 Flash remains weak in average Pearson correlation under both policies, rising only from 0.14 to 0.19. This suggests that speed and interface convenience do not automatically translate into reliable grading behavior.

\begin{figure}[t]
\centering
\includegraphics[width=\linewidth]{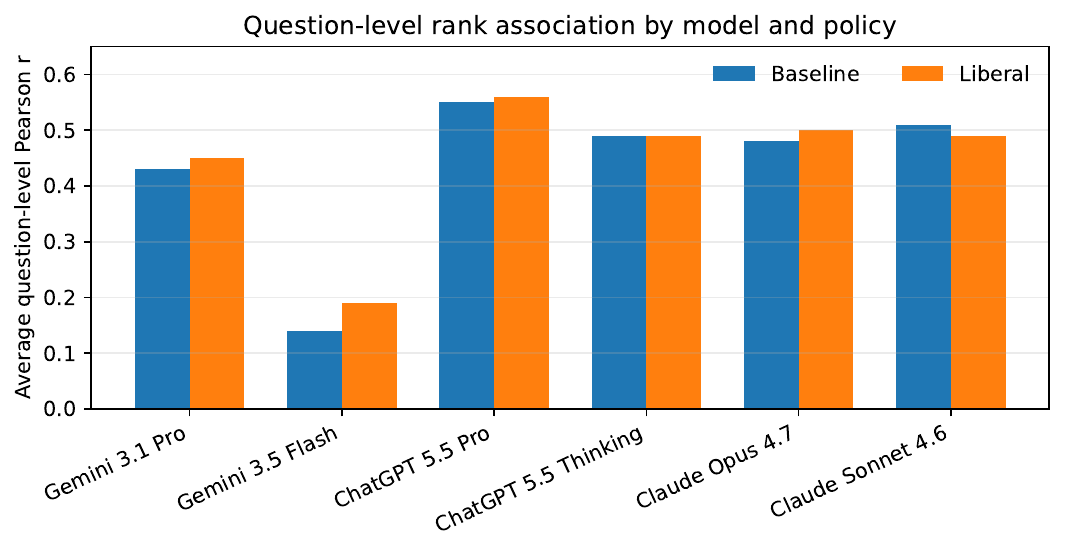}
\caption{Average question-level Pearson correlation by model and grading policy. Higher values indicate stronger linear association with human scores.}
\label{fig:question-pearson}
\end{figure}

\subsubsection{Exact agreement remains low}
Exact agreement is stringent because the model must match the human score exactly. The best average question-level exact agreement is 0.22, achieved by Gemini 3.1 Pro Extended (\Liberal{}) and ChatGPT 5.5 Thinking (\Liberal{}). At the total-score level, exact agreement is near zero for almost all conditions. This does not mean the models are useless, but it does mean that direct release of model-assigned grades would be inappropriate in high-stakes settings. Exact agreement supports a human-in-the-loop deployment model in which LLMs draft scores and feedback, while instructors approve final grades.

\begin{figure}[t]
\centering
\includegraphics[width=\linewidth]{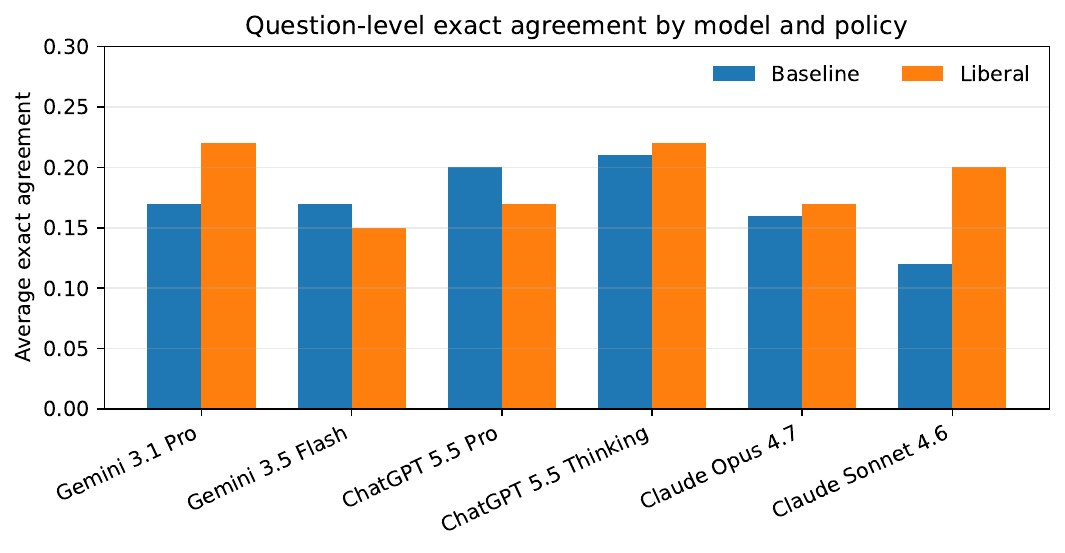}
\caption{Average question-level exact agreement by model and grading policy. Exact agreement is stricter than correlation or error metrics because it requires an exact point match.}
\label{fig:question-exact}
\end{figure}

\subsubsection{Operational usability differs sharply across interfaces}
Performance metrics alone do not capture grading feasibility. In the experiment, ChatGPT was the easiest to use operationally because the paid version completed the grading task in one attempt. Claude Pro Opus 4.7 produced usable grades but consumed enough tokens or session capacity that the interface required a continuation action. Gemini was fast, but the workflow was less reliable on the full dataset: file-not-found errors and inaccurate responses required the dataset to be split into three parts, especially for Gemini 3.5 Flash. These issues matter because grading assistants are useful only if instructors can run them reliably without excessive manual recovery.

\begin{table}[t]
\centering
\scriptsize
\setlength{\tabcolsep}{3pt}
\renewcommand{\arraystretch}{1.08}
\caption{Observed usability notes from the grading workflow. These are practical execution observations, not controlled speed or cost benchmarks.}
\label{tab:usability}
\begin{tabular}{p{2.1cm}p{5.3cm}}
\toprule
Model/interface family & Observed workflow behavior \\
\midrule
ChatGPT paid version & Completed the grading task in one attempt and was the easiest interface to use in this experiment. \\
Claude Pro Opus 4.7 & Reached usage/session limits and required a continuation action; also observed as the slowest condition. \\
Gemini 3.1 Pro Extended & Fast, but experienced file-handling friction, including dataset-file-not-found behavior. \\
Gemini 3.5 Flash & Fast, but did not reliably handle the full dataset in one run; the dataset had to be split into three parts before final grading output was acceptable. \\
\bottomrule
\end{tabular}
\end{table}

\section{Feedback Quality in Mathematics Grading}
\subsection{Why Feedback Quality Requires Separate Design}
A model can assign a numerically accurate score while producing weak feedback. Conversely, fluent feedback can be pedagogically misleading if it is not grounded in the rubric. For mathematics exams, useful feedback should identify the exact reasoning step that earned or lost credit. It should also distinguish among conceptual errors, notation errors, missing justification, arithmetic mistakes, and incomplete solutions.

We therefore treat feedback quality as a structured output requirement rather than as a byproduct of scoring. Table~\ref{tab:feedback-schema} defines the feedback schema used by the proposed assistant. Each generated feedback item should be criterion-level, evidence-based, actionable, and reviewable. In high-stakes contexts, the student-facing message should be released only after instructor approval.

\begin{table}[t]
\centering
\scriptsize
\setlength{\tabcolsep}{3pt}
\renewcommand{\arraystretch}{1.08}
\caption{Feedback-quality schema for mathematics exam grading.}
\label{tab:feedback-schema}
\begin{tabular}{p{2.0cm}p{5.4cm}}
\toprule
Component & Requirement \\
\midrule
Rubric link & Name the rubric criterion or subproblem associated with the feedback. \\
Evidence & Quote or summarize the student's reasoning step that supports the score. \\
Partial-credit logic & Explain why credit was awarded or withheld, including point caps where applicable. \\
Misconception tag & Identify the likely issue, such as missing proof case, incorrect recurrence, invalid set operation, arithmetic error, or unsupported conclusion. \\
Actionable revision & Provide a concise next step without revealing unnecessary solution details. \\
Confidence and escalation & Mark uncertain, ambiguous, operationally incomplete, or borderline cases for instructor review. \\
\bottomrule
\end{tabular}
\end{table}

\subsection{Instructor-Facing Versus Student-Facing Feedback}
The assistant should generate two related but distinct feedback views. The instructor-facing view should include detailed rationales, rubric evidence, confidence, possible alternative scores, and audit flags. This view supports grading review and dispute resolution. The student-facing view should be shorter, constructive, and focused on learning. It should not overstate certainty or expose hidden grading deliberations. Separating these views helps maintain transparency while preserving academic integrity.

\section{Discussion}
\subsection{Implications for LLM-Assisted Mathematics Grading}
The results indicate that LLMs can provide useful first-pass grading assistance for open-ended mathematics exams, but only when embedded in a controlled workflow. Liberal partial-credit prompting substantially improves point calibration in this dataset. This is consistent with the observation that the baseline condition behaved too strictly and sometimes penalized partially correct reasoning more harshly than human graders did.

However, the results also show that no single model-policy pair dominates every metric. ChatGPT 5.5 Thinking (\Liberal{}) is strongest on average question-level MAE and RMSE. Gemini 3.1 Pro Extended (\Liberal{}) is strongest on total-score MAE and total-score RMSE. ChatGPT 5.5 Pro Extended (\Liberal{}) is strongest on average question-level Pearson correlation. Gemini 3.1 Pro Extended (\Baseline{}) is strongest on total-score Pearson correlation. Exact agreement remains low across the board. Thus, instructors should not select a model based only on brand reputation, speed, or one metric. They should validate each model-policy pair on course-specific examples.

\subsection{Why Liberal Partial Credit Helped}
Mathematics grading often involves recognizing incomplete but valid reasoning. A baseline prompt that demands formal rigor can be appropriate for proof-heavy tasks, but it can also be too strict when students demonstrate the correct idea with imperfect notation or incomplete final presentation. The liberal condition improved average MAE for all evaluated models because it made the grading prompt more closely resemble typical human partial-credit behavior. This does not mean liberal grading is always preferable. A liberal prompt can over-award credit if it infers reasoning that is not present. Therefore, liberal policies should be paired with evidence requirements: the model should award credit only when it can identify a student response fragment that supports the criterion.

\begin{figure}[t]
\centering
\includegraphics[width=\linewidth]{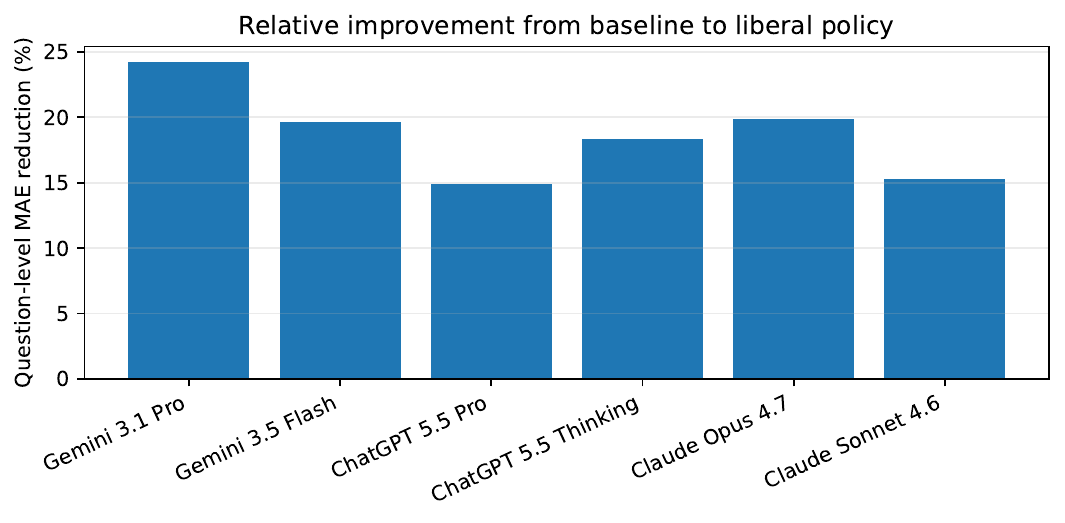}
\caption{Relative reduction in average question-level MAE from the baseline policy to the liberal policy. All evaluated model families show lower question-level MAE under the liberal policy.}
\label{fig:liberal-reduction}
\end{figure}

\subsection{Operational Feasibility Matters}
A grading assistant that produces strong metrics in principle may still be impractical if it cannot process the full dataset reliably. The observed workflow differences are therefore important. ChatGPT completed the grading task in one attempt, which reduced manual coordination. Claude required continuation after reaching usage or session limits, increasing the chance of interrupted state or duplicated work. Gemini was fast but required additional recovery because of file-handling failures and the need to split a large dataset into smaller parts. Dataset splitting can also affect consistency because the model may not see all submissions together, which makes cross-submission calibration harder.

For deployment, operational feasibility should be measured alongside score agreement. Useful measures include number of manual interventions, number of failed file loads, number of continuation actions, run completion rate, average grading time, output parse success rate, and consistency after batching. These measures should be included in future grading benchmarks because they directly affect instructor workload.

\subsection{Why Rubrics Need Calibration}
A static rubric written before seeing student answers may omit common partially correct strategies or unanticipated misconceptions. Human graders often revise their interpretation after grading sample answers. The same principle should apply to LLM-assisted grading. Rubric calibration can be implemented by sampling representative student responses, asking instructors to confirm scores and rationales, and updating the rubric with additional decision rules. In mathematics, calibration is especially important because two solutions can be logically equivalent even when they use different notation or proof structure.

\subsection{Deployment Recommendations}
Based on the empirical results and the proposed workflow, we recommend the following deployment procedure:
\begin{enumerate}
    \item Begin with instructor-authored rubrics and a small calibration set of human-graded examples.
    \item Evaluate at least two prompt policies, including a stricter baseline policy and a liberal partial-credit policy.
    \item Report both error metrics and rank metrics. Do not rely on Pearson correlation alone.
    \item Require criterion-level evidence, confidence, and feedback, not only a final score.
    \item Record operational metrics, including file failures, continuation events, dataset splitting, parse failures, speed, and manual interventions.
    \item Use cross-submission audit to identify inconsistent grading among similar answers.
    \item Keep instructors in control of final grades, especially for flagged cases and high-stakes assessments.
\end{enumerate}

\section{Threats to Validity and Ethical Considerations}
\subsection{Internal Validity}
The study compares model outputs to human reference scores, but human grading itself may contain noise. Future work should include multiple independent human graders, inter-rater reliability, and adjudicated gold scores. Exact model snapshots, decoding settings, interface conditions, and file-processing behavior should also be documented because commercial LLM behavior can change over time.

\subsection{External Validity}
The dataset comes from one discrete mathematics examination. Results may not generalize to other mathematics courses, proof-heavy upper-level courses, handwritten exams, multilingual responses, assessments with diagrams, or different institutional rubrics. Additional datasets are needed to evaluate robustness across course levels and problem types.

\subsection{Construct Validity}
MAE, RMSE, NRMSE, Pearson correlation, and exact agreement measure score alignment, but they do not fully measure feedback quality. This paper defines a feedback-quality schema and deployment workflow, but future studies should collect instructor and student ratings of feedback usefulness, specificity, correctness, tone, and perceived fairness.

\subsection{Operational Validity}
The usability observations in this paper are important but not a controlled benchmark. The claims about ChatGPT completing the task in one attempt, Claude requiring continuation, Gemini requiring dataset splitting, Gemini being fast, and Claude Opus being slow should be interpreted as practical observations from this study, not universal statements about the products. Future work should measure latency, token use, cost, completion rate, and error recovery under controlled conditions.

\subsection{Ethics, Privacy, and Accountability}
LLM-assisted grading must protect student privacy and preserve instructor accountability. Student submissions should be anonymized before model processing, and institutions should ensure compliance with applicable data-governance policies. Instructors should disclose whether AI assistance is used in grading and maintain an audit trail for contested scores. The assistant should never be the sole authority for final grades in high-stakes settings.

\section{Conclusion}
This paper evaluates LLMs as teaching assistants for open-ended mathematics exam grading. Using seven discrete mathematics questions and instructor-assigned reference scores, we evaluate six contemporary LLM configurations under baseline and liberal grading policies. The results show that policy choice is crucial. Liberal partial-credit prompting reduces average question-level and total-score error for every evaluated model family, supporting the decision to add a liberal condition after the baseline prompt proved too strict. ChatGPT 5.5 Thinking (\Liberal{}) achieves the best average question-level point calibration, while Gemini 3.1 Pro Extended (\Liberal{}) achieves the best total-score point calibration. At the same time, the strongest total-score Pearson correlation appears under Gemini 3.1 Pro Extended (\Baseline{}), showing that rank preservation and point agreement must be evaluated separately.

The broader conclusion is that LLMs should be deployed as auditable grading assistants, not autonomous graders. A reliable workflow requires rubric calibration, policy validation, cross-submission audit, structured feedback, operational logging, and human approval. Practical usability is also part of grading quality: file-handling failures, dataset-size limits, session interruptions, token consumption, and speed can determine whether a model is actually useful to instructors. The proposed LLM Evaluation Assistant provides a foundation for future mathematics assessment systems that reduce grading workload while preserving fairness, transparency, and pedagogical value.

\appendices
\section{Prompt Template for Rubric-Grounded Grading}
A production implementation should separate system instructions, rubric content, student response, and output schema. A concise template is shown below.

\begin{quote}
You are assisting an instructor with grading a mathematics exam. Grade only according to the provided rubric. Apply the specified policy: \Baseline{} or \Liberal{}. The \Baseline{} policy requires stricter evidence and complete justification. The \Liberal{} policy may award partial credit for valid mathematical ideas even when notation or presentation is imperfect, but it must not infer reasoning that is not present. Return: (1) final score, (2) criterion-level credit, (3) evidence from the response, (4) missing or incorrect reasoning, (5) student-facing feedback, (6) confidence from 0 to 100, and (7) an escalation flag if the case is ambiguous, inconsistent with similar answers, or affected by file/output problems.
\end{quote}

\section{Instructor Review Checklist}
For each flagged response, the instructor should check whether: (i) the score respects the point caps in the rubric; (ii) similar answers received similar scores; (iii) the rationale cites actual student reasoning; (iv) the feedback is mathematically correct; (v) the feedback is actionable but does not reveal more than intended; (vi) the final grade is recorded with an audit trail; and (vii) the model output was not affected by file-loading errors, continuation truncation, or dataset-splitting inconsistencies.

\section{Full Results Tables}
Because the full result table is large, it is included at the end of the manuscript in landscape orientation.

\onecolumn
\begin{landscape}
\begin{scriptsize}
\setlength{\tabcolsep}{3pt}
\renewcommand{\arraystretch}{1.05}
\centering
\begin{longtable}{p{1.0cm}p{6.3cm}rrrrrr}
\caption{Grading performance of LLM-based graders on the discrete mathematics dataset. Lower MAE, NRMSE, and RMSE are better; higher Pearson and exact agreement are better. The Total row is computed from exam totals.}
\label{tab:llm-grading-performance}\\

\toprule
Question & Method & $N$ & MAE~$(\downarrow)$ & NRMSE~$(\downarrow)$ & RMSE~$(\downarrow)$ & Pearson~$(\uparrow)$ & Exact~$(\uparrow)$ \\
\midrule
\endfirsthead

\multicolumn{8}{c}{\tablename\ \thetable{} -- continued from previous page}\\
\toprule
Question & Method & $N$ & MAE~$(\downarrow)$ & NRMSE~$(\downarrow)$ & RMSE~$(\downarrow)$ & Pearson~$(\uparrow)$ & Exact~$(\uparrow)$ \\
\midrule
\endhead

\midrule
\multicolumn{8}{r}{Continued on next page}\\
\endfoot

\bottomrule
\endlastfoot

\midrule
Q1 & Gemini 3.1 Pro Extended (baseline) & 28 & 1.59 & 0.43 & 2.38 & 0.24 & 0.11 \\
 & Gemini 3.1 Pro Extended (liberal) & 28 & 0.91 & 0.24 & 1.33 & 0.77 & 0.39 \\
 & Gemini 3.5 Flash (baseline) & 28 & 1.91 & 0.51 & 2.80 & 0.06 & 0.25 \\
 & Gemini 3.5 Flash (liberal) & 28 & 1.29 & 0.34 & 1.89 & 0.08 & 0.25 \\
 & ChatGPT 5.5 Pro Extended (baseline) & 27 & 1.09 & 0.29 & 1.60 & 0.52 & 0.33 \\
 & ChatGPT 5.5 Pro Extended (liberal) & 28 & 1.07 & 0.27 & 1.51 & 0.55 & 0.25 \\
 & ChatGPT 5.5 Thinking (baseline) & 28 & 1.16 & 0.30 & 1.68 & 0.46 & 0.32 \\
 & ChatGPT 5.5 Thinking (liberal) & 28 & 0.95 & 0.27 & 1.51 & 0.44 & 0.43 \\
 & Claude Pro Opus 4.7 (baseline) & 28 & 1.34 & 0.31 & 1.69 & 0.60 & 0.21 \\
 & Claude Pro Opus 4.7 (liberal) & 28 & 0.98 & 0.21 & 1.18 & 0.66 & 0.21 \\
 & Claude Sonnet 4.6 (baseline) & 28 & 0.91 & 0.21 & 1.18 & 0.69 & 0.25 \\
 & Claude Sonnet 4.6 (liberal) & 28 & 0.91 & 0.21 & 1.15 & 0.65 & 0.25 \\

\midrule
Q2 & Gemini 3.1 Pro Extended (baseline) & 28 & 2.66 & 0.33 & 3.34 & 0.42 & 0.11 \\
 & Gemini 3.1 Pro Extended (liberal) & 28 & 1.80 & 0.26 & 2.61 & 0.56 & 0.25 \\
 & Gemini 3.5 Flash (baseline) & 28 & 2.86 & 0.36 & 3.59 & 0.21 & 0.11 \\
 & Gemini 3.5 Flash (liberal) & 28 & 2.18 & 0.30 & 3.00 & 0.24 & 0.21 \\
 & ChatGPT 5.5 Pro Extended (baseline) & 27 & 2.76 & 0.33 & 3.25 & 0.70 & 0.11 \\
 & ChatGPT 5.5 Pro Extended (liberal) & 28 & 2.55 & 0.32 & 3.16 & 0.60 & 0.11 \\
 & ChatGPT 5.5 Thinking (baseline) & 28 & 3.45 & 0.40 & 4.03 & 0.59 & 0.14 \\
 & ChatGPT 5.5 Thinking (liberal) & 28 & 2.73 & 0.32 & 3.25 & 0.61 & 0.07 \\
 & Claude Pro Opus 4.7 (baseline) & 28 & 2.95 & 0.35 & 3.48 & 0.54 & 0.07 \\
 & Claude Pro Opus 4.7 (liberal) & 28 & 2.27 & 0.27 & 2.71 & 0.56 & 0.07 \\
 & Claude Sonnet 4.6 (baseline) & 28 & 3.12 & 0.37 & 3.66 & 0.52 & 0.07 \\
 & Claude Sonnet 4.6 (liberal) & 28 & 2.59 & 0.30 & 3.03 & 0.54 & 0.11 \\

\midrule
Q3 & Gemini 3.1 Pro Extended (baseline) & 28 & 3.30 & 0.67 & 4.00 & 0.43 & 0.11 \\
 & Gemini 3.1 Pro Extended (liberal) & 28 & 2.16 & 0.47 & 2.84 & 0.09 & 0.14 \\
 & Gemini 3.5 Flash (baseline) & 28 & 2.38 & 0.50 & 3.01 & 0.08 & 0.14 \\
 & Gemini 3.5 Flash (liberal) & 28 & 1.88 & 0.41 & 2.43 & 0.17 & 0.00 \\
 & ChatGPT 5.5 Pro Extended (baseline) & 27 & 2.54 & 0.54 & 3.25 & 0.41 & 0.11 \\
 & ChatGPT 5.5 Pro Extended (liberal) & 28 & 2.21 & 0.46 & 2.75 & 0.26 & 0.07 \\
 & ChatGPT 5.5 Thinking (baseline) & 28 & 2.80 & 0.59 & 3.51 & 0.11 & 0.14 \\
 & ChatGPT 5.5 Thinking (liberal) & 28 & 2.23 & 0.49 & 2.94 & 0.12 & 0.25 \\
 & Claude Pro Opus 4.7 (baseline) & 28 & 2.66 & 0.56 & 3.36 & 0.26 & 0.18 \\
 & Claude Pro Opus 4.7 (liberal) & 28 & 2.16 & 0.45 & 2.72 & 0.26 & 0.07 \\
 & Claude Sonnet 4.6 (baseline) & 28 & 2.80 & 0.58 & 3.45 & 0.18 & 0.07 \\
 & Claude Sonnet 4.6 (liberal) & 28 & 2.34 & 0.53 & 3.17 & 0.01 & 0.21 \\

\midrule
Q4 & Gemini 3.1 Pro Extended (baseline) & 28 & 2.86 & 0.39 & 3.91 & 0.36 & 0.25 \\
 & Gemini 3.1 Pro Extended (liberal) & 28 & 2.43 & 0.32 & 3.23 & 0.47 & 0.14 \\
 & Gemini 3.5 Flash (baseline) & 28 & 3.71 & 0.43 & 4.34 & 0.01 & 0.04 \\
 & Gemini 3.5 Flash (liberal) & 28 & 3.07 & 0.38 & 3.80 & 0.02 & 0.07 \\
 & ChatGPT 5.5 Pro Extended (baseline) & 27 & 3.33 & 0.43 & 4.29 & 0.26 & 0.15 \\
 & ChatGPT 5.5 Pro Extended (liberal) & 28 & 2.38 & 0.33 & 3.29 & 0.53 & 0.11 \\
 & ChatGPT 5.5 Thinking (baseline) & 28 & 2.93 & 0.39 & 3.89 & 0.49 & 0.25 \\
 & ChatGPT 5.5 Thinking (liberal) & 28 & 2.54 & 0.33 & 3.35 & 0.46 & 0.11 \\
 & Claude Pro Opus 4.7 (baseline) & 28 & 3.36 & 0.42 & 4.24 & 0.36 & 0.18 \\
 & Claude Pro Opus 4.7 (liberal) & 28 & 2.79 & 0.36 & 3.65 & 0.41 & 0.18 \\
 & Claude Sonnet 4.6 (baseline) & 28 & 3.54 & 0.43 & 4.34 & 0.21 & 0.14 \\
 & Claude Sonnet 4.6 (liberal) & 28 & 3.14 & 0.40 & 4.01 & 0.23 & 0.18 \\

\midrule
Q5 & Gemini 3.1 Pro Extended (baseline) & 28 & 2.50 & 0.58 & 3.48 & 0.52 & 0.29 \\
 & Gemini 3.1 Pro Extended (liberal) & 28 & 1.18 & 0.29 & 1.73 & 0.82 & 0.32 \\
 & Gemini 3.5 Flash (baseline) & 28 & 1.54 & 0.41 & 2.48 & 0.43 & 0.36 \\
 & Gemini 3.5 Flash (liberal) & 28 & 1.23 & 0.34 & 2.02 & 0.55 & 0.32 \\
 & ChatGPT 5.5 Pro Extended (baseline) & 27 & 2.41 & 0.57 & 3.40 & 0.71 & 0.26 \\
 & ChatGPT 5.5 Pro Extended (liberal) & 28 & 1.89 & 0.47 & 2.80 & 0.75 & 0.29 \\
 & ChatGPT 5.5 Thinking (baseline) & 28 & 1.86 & 0.45 & 2.72 & 0.67 & 0.21 \\
 & ChatGPT 5.5 Thinking (liberal) & 28 & 1.50 & 0.39 & 2.31 & 0.67 & 0.29 \\
 & Claude Pro Opus 4.7 (baseline) & 28 & 2.57 & 0.56 & 3.34 & 0.70 & 0.11 \\
 & Claude Pro Opus 4.7 (liberal) & 28 & 1.82 & 0.43 & 2.60 & 0.72 & 0.29 \\
 & Claude Sonnet 4.6 (baseline) & 28 & 2.71 & 0.53 & 3.20 & 0.81 & 0.07 \\
 & Claude Sonnet 4.6 (liberal) & 28 & 2.25 & 0.46 & 2.75 & 0.84 & 0.14 \\

\midrule
Q6 & Gemini 3.1 Pro Extended (baseline) & 28 & 1.82 & 0.61 & 2.45 & 0.63 & 0.14 \\
 & Gemini 3.1 Pro Extended (liberal) & 28 & 1.73 & 0.56 & 2.24 & 0.38 & 0.18 \\
 & Gemini 3.5 Flash (baseline) & 28 & 1.80 & 0.60 & 2.42 & 0.28 & 0.14 \\
 & Gemini 3.5 Flash (liberal) & 28 & 1.61 & 0.50 & 2.00 & 0.24 & 0.14 \\
 & ChatGPT 5.5 Pro Extended (baseline) & 27 & 1.28 & 0.44 & 1.77 & 0.61 & 0.22 \\
 & ChatGPT 5.5 Pro Extended (liberal) & 28 & 1.25 & 0.46 & 1.85 & 0.61 & 0.21 \\
 & ChatGPT 5.5 Thinking (baseline) & 28 & 1.55 & 0.52 & 2.09 & 0.51 & 0.21 \\
 & ChatGPT 5.5 Thinking (liberal) & 28 & 1.27 & 0.43 & 1.70 & 0.50 & 0.18 \\
 & Claude Pro Opus 4.7 (baseline) & 28 & 2.34 & 0.75 & 3.00 & 0.49 & 0.11 \\
 & Claude Pro Opus 4.7 (liberal) & 28 & 1.84 & 0.60 & 2.41 & 0.46 & 0.14 \\
 & Claude Sonnet 4.6 (baseline) & 28 & 1.77 & 0.58 & 2.32 & 0.62 & 0.11 \\
 & Claude Sonnet 4.6 (liberal) & 28 & 1.45 & 0.52 & 2.07 & 0.57 & 0.25 \\

\midrule
Q7 & Gemini 3.1 Pro Extended (baseline) & 28 & 2.61 & 0.38 & 3.81 & 0.42 & 0.18 \\
 & Gemini 3.1 Pro Extended (liberal) & 28 & 2.93 & 0.40 & 4.04 & 0.08 & 0.11 \\
 & Gemini 3.5 Flash (baseline) & 28 & 3.30 & 0.44 & 4.40 & -0.07 & 0.18 \\
 & Gemini 3.5 Flash (liberal) & 28 & 2.84 & 0.40 & 3.97 & 0.02 & 0.07 \\
 & ChatGPT 5.5 Pro Extended (baseline) & 27 & 2.11 & 0.30 & 3.04 & 0.65 & 0.19 \\
 & ChatGPT 5.5 Pro Extended (liberal) & 28 & 1.86 & 0.28 & 2.76 & 0.61 & 0.18 \\
 & ChatGPT 5.5 Thinking (baseline) & 28 & 2.25 & 0.30 & 2.99 & 0.62 & 0.18 \\
 & ChatGPT 5.5 Thinking (liberal) & 28 & 1.86 & 0.26 & 2.65 & 0.64 & 0.25 \\
 & Claude Pro Opus 4.7 (baseline) & 28 & 2.43 & 0.33 & 3.34 & 0.40 & 0.25 \\
 & Claude Pro Opus 4.7 (liberal) & 28 & 2.29 & 0.32 & 3.17 & 0.45 & 0.21 \\
 & Claude Sonnet 4.6 (baseline) & 27 & 2.59 & 0.33 & 3.29 & 0.55 & 0.15 \\
 & Claude Sonnet 4.6 (liberal) & 27 & 2.11 & 0.28 & 2.84 & 0.57 & 0.26 \\

\midrule
Total & Gemini 3.1 Pro Extended (baseline) & 28 & 12.66 & 0.51 & 15.24 & 0.58 & 0.00 \\
 & Gemini 3.1 Pro Extended (liberal) & 28 & 8.00 & 0.36 & 10.66 & 0.34 & 0.04 \\
 & Gemini 3.5 Flash (baseline) & 28 & 12.46 & 0.56 & 16.85 & 0.01 & 0.04 \\
 & Gemini 3.5 Flash (liberal) & 28 & 8.16 & 0.36 & 10.88 & 0.20 & 0.00 \\
 & ChatGPT 5.5 Pro Extended (baseline) & 27 & 14.70 & 0.68 & 17.59 & 0.38 & 0.00 \\
 & ChatGPT 5.5 Pro Extended (liberal) & 28 & 11.79 & 0.49 & 14.61 & 0.41 & 0.04 \\
 & ChatGPT 5.5 Thinking (baseline) & 28 & 15.96 & 0.59 & 17.84 & 0.46 & 0.00 \\
 & ChatGPT 5.5 Thinking (liberal) & 28 & 11.25 & 0.46 & 13.68 & 0.45 & 0.00 \\
 & Claude Pro Opus 4.7 (baseline) & 28 & 16.29 & 0.62 & 18.71 & 0.48 & 0.00 \\
 & Claude Pro Opus 4.7 (liberal) & 28 & 11.14 & 0.45 & 13.59 & 0.46 & 0.00 \\
 & Claude Sonnet 4.6 (baseline) & 28 & 18.39 & 0.68 & 20.27 & 0.46 & 0.00 \\
 & Claude Sonnet 4.6 (liberal) & 28 & 13.75 & 0.54 & 16.21 & 0.44 & 0.00 \\

\end{longtable}
\end{scriptsize}
\end{landscape}


\begin{table}[t]
\centering
\scriptsize
\setlength{\tabcolsep}{3pt}
\renewcommand{\arraystretch}{1.05}
\caption{Overall grading performance averaged over Q1--Q7 on the discrete mathematics dataset.}
\label{tab:llm-grading-averages}
\begin{tabular}{p{6.3cm}rrrrrr}
\toprule
Method & Avg. $N$ & MAE~$(\downarrow)$ & NRMSE~$(\downarrow)$ & RMSE~$(\downarrow)$ & Pearson~$(\uparrow)$ & Exact~$(\uparrow)$ \\
\midrule
Gemini 3.1 Pro Extended (baseline) & 28 & 2.48 & 0.49 & 3.34 & 0.43 & 0.17 \\
Gemini 3.1 Pro Extended (liberal) & 28 & 1.88 & 0.36 & 2.57 & 0.45 & 0.22 \\
Gemini 3.5 Flash (baseline) & 28 & 2.50 & 0.47 & 3.29 & 0.14 & 0.17 \\
Gemini 3.5 Flash (liberal) & 28 & 2.01 & 0.38 & 2.73 & 0.19 & 0.15 \\
ChatGPT 5.5 Pro Extended (baseline) & 27 & 2.22 & 0.41 & 2.94 & 0.55 & 0.20 \\
ChatGPT 5.5 Pro Extended (liberal) & 28 & 1.89 & 0.37 & 2.59 & 0.56 & 0.17 \\
ChatGPT 5.5 Thinking (baseline) & 28 & 2.29 & 0.42 & 2.99 & 0.49 & 0.21 \\
ChatGPT 5.5 Thinking (liberal) & 28 & 1.87 & 0.36 & 2.53 & 0.49 & 0.22 \\
Claude Pro Opus 4.7 (baseline) & 28 & 2.52 & 0.47 & 3.21 & 0.48 & 0.16 \\
Claude Pro Opus 4.7 (liberal) & 28 & 2.02 & 0.38 & 2.63 & 0.50 & 0.17 \\
Claude Sonnet 4.6 (baseline) & 27.9 & 2.49 & 0.43 & 3.06 & 0.51 & 0.12 \\
Claude Sonnet 4.6 (liberal) & 27.9 & 2.11 & 0.39 & 2.72 & 0.49 & 0.20 \\
\bottomrule
\end{tabular}
\end{table}
\twocolumn

\balance
\bibliographystyle{IEEEtran}
\bibliography{references}

@article{wang2026large,
  title={Large language models for education: A survey and outlook},
  author={Wang, Shen and Xu, Tianlong and Li, Hang and Zhang, Chaoli and Liang, Joleen and Tang, Jiliang and Yu, Philip S and Wen, Qingsong},
  journal={IEEE Signal Processing Magazine},
  volume={42},
  number={6},
  pages={51--63},
  year={2026},
  publisher={IEEE}
}

@article{chu2024llm,
  title={A llm-powered automatic grading framework with human-level guidelines optimization},
  author={Chu, Yucheng and Li, Hang and Yang, Kaiqi and Shomer, Harry and Liu, Hui and Copur-Gencturk, Yasemin and Tang, Jiliang},
  journal={arXiv preprint arXiv:2410.02165},
  year={2024}
}

@inproceedings{satcharattanachot2025automated,
  title={Automated Grading Approach for Open-Ended STEM Answers using LLM},
  author={Satcharattanachot, Piyanat and Usanavasin, Sasiporn},
  booktitle={2025 20th International Joint Symposium on Artificial Intelligence and Natural Language Processing (iSAI-NLP)},
  pages={1--6},
  year={2025},
  organization={IEEE}
}

@article{alkafaween2025automating,
  title={Automating autograding: Large language models as test suite generators for introductory programming},
  author={Alkafaween, Umar and Albluwi, Ibrahim and Denny, Paul},
  journal={Journal of Computer Assisted Learning},
  volume={41},
  number={1},
  pages={e13100},
  year={2025},
  publisher={Wiley Online Library}
}

@inproceedings{chu2025llm,
  title={LLM-based Automated Grading with Human-in-the-Loop},
  author={Chu, Yucheng and Li, Hang and Yang, Kaiqi and Copur-Gencturk, Yasemin and Tang, Jiliang},
  booktitle={2025 IEEE International Conference on Teaching, Assessment, and Learning for Engineering (TALE)},
  pages={1--8},
  year={2025},
  organization={IEEE}
}

@article{lee2026personalized,
  title={Personalized Auto-Grading and Feedback System for Constructive Geometry Tasks Using Large Language Models on an Online Math Platform},
  author={Lee, Yong Oh and Bang, Byeonghun and Lee, Joohyun and Oh, Sejun},
  journal={IEEE Access},
  year={2026},
  publisher={IEEE}
}

@article{schneider2023towards,
  title={Towards llm-based autograding for short textual answers},
  author={Schneider, Johannes and Schenk, Bernd and Niklaus, Christina},
  journal={arXiv preprint arXiv:2309.11508},
  year={2023}
}

@article{impey2025using,
  title={Using large language models for automated grading of student writing about science},
  author={Impey, Chris and Wenger, Matthew and Garuda, Nikhil and Golchin, Shahriar and Stamer, Sarah},
  journal={International Journal of Artificial Intelligence in Education},
  volume={35},
  number={4},
  pages={1825--1859},
  year={2025},
  publisher={Springer}
}

@inproceedings{lee2024college,
  title={College exam grader using LLM AI models},
  author={Lee, Jung X and Song, Yeong-Tae},
  booktitle={2024 IEEE/ACIS 27th International Conference on Software Engineering, Artificial Intelligence, Networking and Parallel/Distributed Computing (SNPD)},
  pages={282--289},
  year={2024},
  organization={IEEE}
}

@article{chu2025llmagentforeducation,
  title={Llm agents for education: Advances and applications},
  author={Chu, Zhendong and Wang, Shen and Xie, Jian and Zhu, Tinghui and Yan, Yibo and Ye, Jinheng and Zhong, Aoxiao and Hu, Xuming and Liang, Jing and Yu, Philip S and others},
  journal={arXiv preprint arXiv:2503.11733},
  volume={2},
  year={2025},
  publisher={Mar}
}

@inproceedings{xie2025grade,
  title={Grade like a human: Rethinking automated assessment with large language models},
  author={Xie, Wenjing and Niu, Juxin and Xue, Chun Jason and Guan, Nan},
  booktitle={Proceedings of the International Conference on Research in Adaptive and Convergent Systems},
  pages={1--8},
  year={2025}
}

@inproceedings{ferreira2025automatic,
  title={Automatic short answer grading in the LLM era: Does GPT-4 with prompt engineering beat traditional models?},
  author={Ferreira Mello, Rafael and Pereira Junior, Cleon and Rodrigues, Luiz and Pereira, Filipe Dwan and Cabral, Luciano and Costa, Newarney and Ramalho, Geber and Gasevic, Dragan},
  booktitle={Proceedings of the 15th international learning analytics and knowledge conference},
  pages={93--103},
  year={2025}
}

@article{baral2024automated,
  title={Automated feedback in math education: A comparative analysis of LLMs for open-ended responses},
  author={Baral, Sami and Worden, Eamon and Lim, Wen-Chiang and Luo, Zhuang and Santorelli, Christopher and Gurung, Ashish and Heffernan, Neil},
  journal={arXiv preprint arXiv:2411.08910},
  year={2024}
}

\end{document}